\newcommand{\printstyle}{reprint}

\documentclass[aps,prl,superscriptaddress,floatfix,\printstyle ,nofootinbib]{revtex4-2}

\usepackage{color}

\usepackage[caption=false]{subfig}
\usepackage{graphicx,booktabs,array}
\usepackage{multirow}
\usepackage{tabularx} 
\DeclareGraphicsExtensions{.pdf,.png,.jpg,.eps,.ps}
\usepackage{epsfig}
\usepackage{epstopdf}
\usepackage{amssymb}
\usepackage{amsmath}
\usepackage{amsfonts}
\usepackage{mathrsfs}
\usepackage{amsthm}
\usepackage{float}
\usepackage{amsbsy}
\usepackage{bm}
\usepackage{grffile}
\usepackage{hyperref}
\hypersetup{
    colorlinks = true,
    citecolor = blue,
    urlcolor = blue,
    linkcolor = blue,
}
\usepackage{courier}
\usepackage{upgreek}
\usepackage{xspace}
\usepackage{xcolor}
\usepackage{calc} 
\usepackage{soul} 
\usepackage[export]{adjustbox} 

\newcommand{\ie}{\textit{i}.\textit{e}. }
\newcommand{\eg}{\textit{e}.\textit{g}. }

\newcommand{\vs}{v_S}

\newcommand{\Epar}{E_\parallel}

\newcommand{\kpar}{k_\parallel}

\newcommand{\neo}{n_e}

\newcommand{\vpar}{v_\parallel}

\newcommand{\betae}{\beta_e}

\newcommand{\npar}{n_\parallel}

\newcommand{\Te}{T_e}
\newcommand{\Ti}{T_i}

\renewcommand{\deg}{^\circ}

\newcommand{\That}{\hat{T}_e}

\newcommand{\gam}{\gamma}
\newcommand{\gxy}{\gam^2}
\newcommand{\cpsd}[2]{\smavg{#1^*#2}}

\newcommand{\dte}{\delta \That}

\newcommand{\ptsolver}{PT\_SOLVER\xspace}

\newcommand{\Teo}{T_{e0}}

\newcommand{\Ip}{I_p}
\newcommand{\Bt}{B_T}
\newcommand{\omod}{\omega_m}

\newcommand{\mcubed}{\text{ m}^3}

\newcommand{\smavg}[1]{\langle #1 \rangle}
\newcommand{\smabs}[1]{\lvert #1 \rvert}
\newcommand{\smIm}[1]{\text{Im}[#1]}

\newcommand{\ten}[1]{\cdot 10^{#1}}

\newcommand{\defined}{\equiv}
\newcommand{\like}{\sim}

\newcommand{\approptoinn}[2]{\mathrel{\vcenter{
  \offinterlineskip\halign{\hfil$##$\cr
    #1\propto\cr\noalign{\kern2pt}#1\sim\cr\noalign{\kern-2pt}}}}}

\newcommand{\figref}[1]{Fig.\xspace\ref{#1}}
\renewcommand{\eqref}[1]{Eq.\xspace\ref{#1}}

\newcommand{\citeref}[1]{Ref.\xspace\onlinecite{#1}}

\newcommand{\myname}{J.B. Lestz}

\newcommand{\Shawn}{S.X. Tang}

\newcommand{\Bart}{B. Van Compernolle}
\newcommand{\Andrea}{A.M. Garofalo}
\newcommand{\Craig}{C.C. Petty}
\newcommand{\Bob}{R.I. Pinsker}
\newcommand{\AlexDupuy}{A. Dupuy}

\newcommand{\Grant}{G. Rutherford}
\newcommand{\Miklos}{M. Porkolab}
\newcommand{\MikeRoss}{M.P. Ross}
\newcommand{\CharlesMoeller}{C.P. Moeller}
\newcommand{\Levi}{L. McAllister}
\newcommand{\AntonioTorrezan}{A. Torrezan}

\newcommand{\UCIshort}{University of California, Irvine, CA 92697, USA}

\newcommand{\GA}{General Atomics, San Diego, CA, 92121, USA}
\newcommand{\MIT}{Massachusetts Institute of Technology, Cambridge, MA, 02139, USA}
\newcommand{\FarTech}{Far-Tech, Inc, San Diego, CA, 92121, USA}

\definecolor{darkgreen}{rgb}{0,0.5,0}

\newcommand{\gadisclaimer}{This report was prepared as an account of work sponsored by an agency of the United States Government. Neither the United States Government nor any agency thereof, nor any of their employees, makes any warranty, express or implied, or assumes any legal liability or responsibility for the accuracy, completeness, or usefulness of any information, apparatus, product, or process disclosed, or represents that its use would not infringe privately owned rights. Reference herein to any specific commercial product, process, or service by trade name, trademark, manufacturer, or otherwise does not necessarily constitute or imply its endorsement, recommendation, or favoring by the United States Government or any agency thereof. The views and opinions of authors expressed herein do not necessarily state or reflect those of the United States Government or any agency thereof.}
\renewcommand{\vs}{{\xspace}versus\xspace}

\newcommand{\sectext}[1]{\textit{#1}\textemdash\xspace}

\newcommand{\vshrink}{\vspace*{-0.25cm}}

\newcommand{\authsymb}{$^\S$}
\newcommand{\authnote}{These authors contributed equally to this work.}

\usepackage[caption=false]{subfig}

\definecolor{red_new}{RGB}{146,0,0}

\begin{document}

\title{First Experimental Evidence of Helicon Current Drive}

\author{\myname\authsymb}
\email{lestzj@fusion.gat.com}
\affiliation{\GA}
\affiliation{\UCIshort}
\author{\Bob\authsymb}
\email{pinsker@far-tech.com}
\affiliation{\GA}
\affiliation{\FarTech}
\author{\Bart\authsymb}
\email{vancompernolle@fusion.gat.com}
\affiliation{\GA}
\author{\Shawn}
\affiliation{\GA}
\author{\AlexDupuy}
\affiliation{\GA}
\author{\Andrea}
\affiliation{\GA}
\author{\Levi}
\affiliation{\GA}
\author{\CharlesMoeller}
\affiliation{\GA}
\author{\Craig}
\affiliation{\GA}
\author{\Miklos}
\affiliation{\MIT}
\author{\MikeRoss}
\affiliation{\GA}
\author{\Grant}
\affiliation{\MIT}
\author{\AntonioTorrezan}
\affiliation{\GA}
\date{\today}
\begin{abstract}

Helicon current drive is an attractive solution for driving current to sustain steady state tokamak operation in reactor conditions. Dedicated DIII-D experiments have been conducted with a MW-level helicon system and successfully demonstrated core power deposition and current drive with helicon waves launched via a traveling wave antenna. The profile of the measured electron temperature response to helicon power injection is in good agreement with time-dependent integrated modeling that incorporates ray tracing and the effects of thermal transport simultaneously. When the helicon power is injected continuously to drive co-$\Ip$ current, the reconstructed safety factor profile flattens significantly faster and sawteeth are triggered earlier than in comparison shots where the helicon is replaced by a comparable amount of electron cyclotron heating. Calculation of the helicon-driven current profile yields a peaked profile in the core, consistent with the observed power deposition profile and in good agreement with ray tracing predictions. Taken together, these experimental results represent strong evidence for the first definitive observation of auxiliary current drive due to helicon waves on any device. 

\end{abstract}
\maketitle

\def\thefootnote{\authsymb}\footnotetext{\authnote}\def\thefootnote{\arabic{footnote}}

\sectext{Introduction}
Steady state tokamak reactor designs require efficient sources of off-axis, non-inductive current drive in order to supplement the bootstrap current and sustain high performing scenarios \cite{Jardin1997FED,ITER1999NFhcd,Gormezano2007NF,Luce2011POP,Kikuchi2012RMP,Buttery2021NF,Litaudon2024NF}. Fast waves in the lower hybrid range of frequencies (LHRF), also known as helicon waves, are an attractive candidate to provide auxiliary current drive in a reactor \cite{Vdovin2013PPR,Prater2014NF}. Helicon waves are resonantly absorbed via Landau damping on electrons with parallel velocities $\vpar = \omega/\kpar$ \cite{Landau1946,Stix1975NF}. A net current is driven by launching a parallel wave spectrum with strong directivity, \ie dominantly accelerating electrons in one toroidal direction \cite{Chiu1989NF}. At sufficiently high volume-averaged electron pressure $\smavg{\betae} = 2\mu_0 \smavg{\neo \Te/B^2} \gtrsim 2\%$, such as those anticipated in future reactors, helicon waves are predicted to be fully absorbed at mid-radius on a single pass \cite{Prater2014NF}, with higher current drive efficiency than achievable with off-axis electron cyclotron current drive and without the accessibility limitations of lower hybrid current drive (LHCD), which can not penetrate deep into the plasma at reactor-relevant densities \cite{Pinsker2015POP}. Hence, helicon waves hold promise as a complementary, reactor-relevant current drive actuator alongside other radio frequency wave schemes. 

Despite efforts made by multiple groups, helicon current drive had not yet been experimentally demonstrated. Previous helicon current drive experiments performed on other tokamaks were conducted in a density regime where both slow waves (commonly referred to as lower hybrid waves) and fast waves could propagate in large regions of the plasma. Consequently, current drive observations from launching fast waves in JIPP-TIIU \cite{Ohkubo1986PRL}, PLT \cite{Pinsker1986thesis,Pinsker1987AIP}, and JFT-2M \cite{Uesugi1987AIP} were all explained as resulting from unintended mode conversion to slow waves, as the observed current drive efficiencies and density dependence were nearly identical to those resulting from LHCD \cite{Pinsker1994AIP}. Historically, technological limitations shifted the investigation of fast wave current drive (FWCD) from the LHRF to the ion cyclotron range of frequencies (ICRF). FWCD in the ICRF was subsequently demonstrated in JIPP-TIIU \cite{Ando1986NF}, DIII-D \cite{Prater1993PPCF,Petty1995NF,deGrassie1996AIP,Petty1997AIP,Petty1999NF,Petty2001PPCF}, Tore Supra \cite{Saoutic1994PPCF}, TFTR \cite{Rogers1996AIP,Majeski1996POP}, and NSTX \cite{Hosea2008POP,Phillips2009NF,Taylor2012POP} experiments that could conclusively attribute the current to fast waves by operating above the LHCD density limit. The maturation of FWCD in the ICRF motivated further development of helicon current drive, which projects more favorably to reactors since the higher wave frequency increases the rate of fast wave absorption on electrons while rendering undesirable ion cyclotron damping negligible. In addition to the DIII-D results presented here, helicon current drive has been considered for many modern \cite{Li2020PLA,Zi2025FED,Li2020PPCF,Tang2025AIP,Du2025PPCF,Du2025PPCF} and next-step tokamaks \cite{Koch2011AIP,Kim2015NF,Mikkelsen2018NF,Li2021FED,Wu2023NF}, including KSTAR where a helicon traveling wave antenna (TWA) has also been installed and commissioning has begun \cite{Wang2017NF,Wi2023FED,Kim2025RF}. In this Letter, we report the first definitive experimental observation of helicon current drive on any device, operating in a regime where indirect LHCD can be entirely excluded, and find strong agreement between measurements and predictions from modeling.

\newcommand{\epsheight}{4.5cm}
\newcommand{\jpgheight}{4cm}
\begin{figure*}[tb]
\hspace*{0.2cm}
\subfloat{\includegraphics[height = \epsheight]{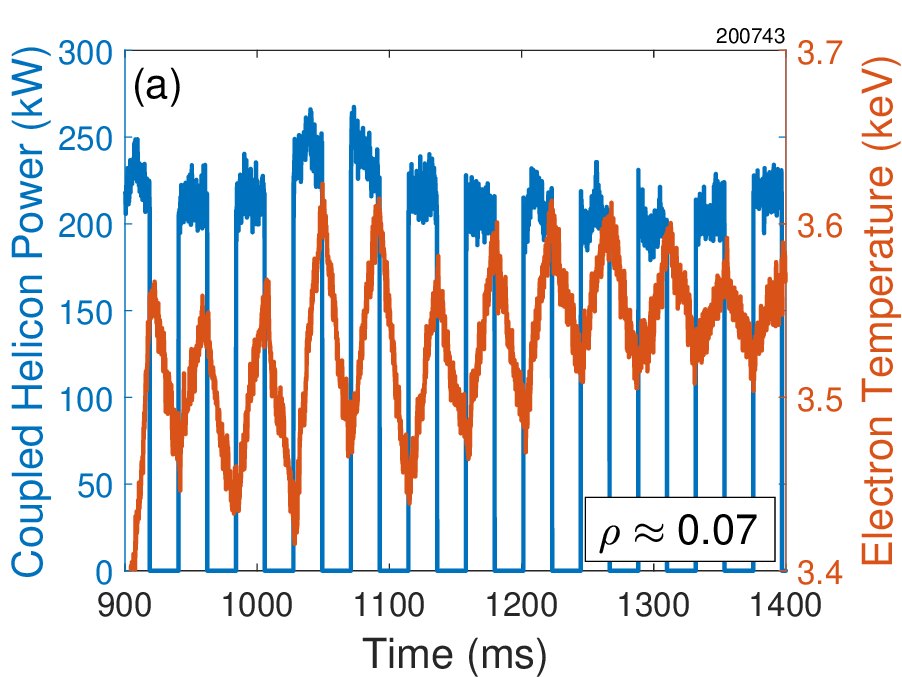}\label{fig:temod}} \hspace*{0.2cm}
\subfloat{\includegraphics[height = \epsheight]{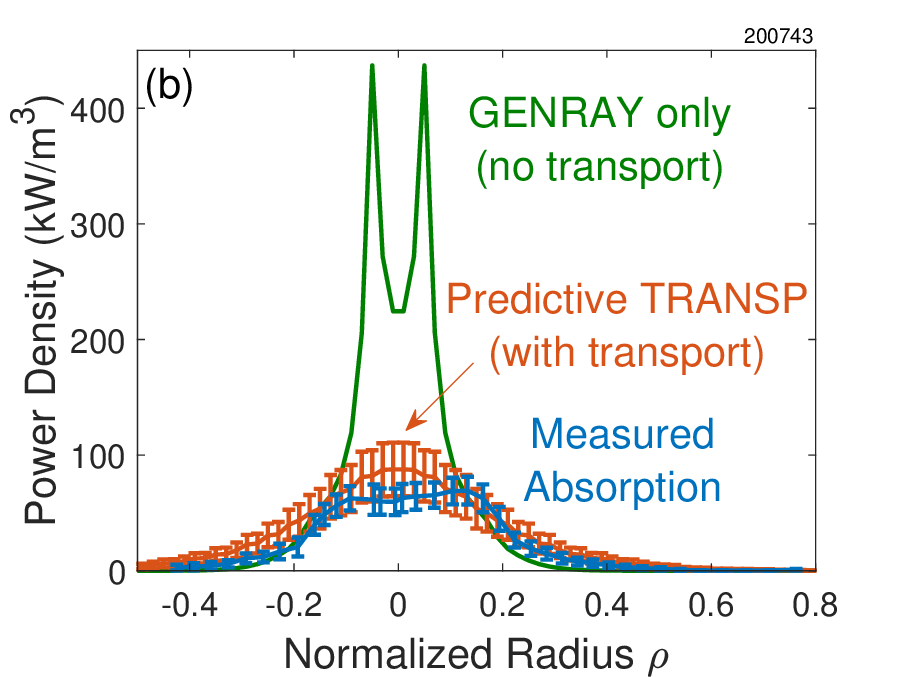}\label{fig:powdep}} \hspace*{-0.4cm}
\subfloat{\includegraphics[height = \epsheight]{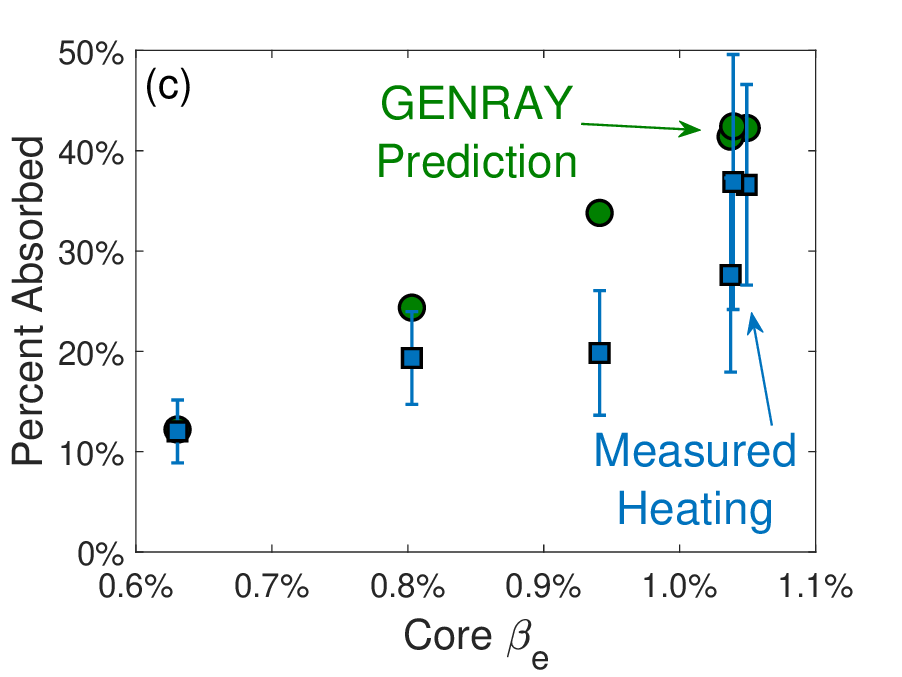}\label{fig:betascan}} \hfill \\ \vshrink
\caption{(a) Observed $\Te$ response to modulated helicon power. (b) Power deposition measured experimentally (blue), predicted by GENRAY (green), and inferred from predictive TRANSP simulations including transport (orange). Negative $\rho$ values correspond to high field side data, whereas the TRANSP and GENRAY calculations are inherently symmetric. (c) Absorbed fraction of coupled helicon power as observed experimentally (blue) and predicted by GENRAY on the first pass (green) as a function of $\betae$ (volume averaged within $\rho < 0.2$). Data from a subset of DIII-D discharges 200737 -- 200743.}
\label{fig:heating}
\end{figure*}

\sectext{Experimental Apparatus}
Experiments were performed with a high power helicon system on the DIII-D tokamak that is designed to launch helicon waves at 476 MHz with $\npar = \kpar c/\omega \approx 3$. DIII-D is a medium sized tokamak with a major radius of approximately 1.7 m and an aspect ratio around 2.7 \cite{Holcomb2024NF}. A 30-module comb-line TWA spans 1.5 m toroidally and 0.2 m poloidally, with the antenna located about 35$\deg$ poloidally above the midplane on the low field side (shown in the End Matter) \cite{Torreblanca2019FED,VanCompernolle2021NF}. A 1.2 MW klystron can feed power to either end of the antenna in order to inject helicon waves in either toroidal direction. Up to 700 kW of helicon power has been coupled to the plasma, to date. Power flows toroidally through the antenna modules via mutual inductance, with low resistive losses \cite{Moeller1994AIP,VanCompernolle2021NF,Squire2023AIP}. Robust load resilience to rapid changes to edge conditions, \eg due to edge localized modes in H-mode, and strong coupling to the plasma have been previously verified \cite{Pinsker2024NF}. Each module of the TWA is tilted approximately 14$\deg$ from the horizontal in order to align with the field lines at the face of the antenna to suppress direct excitation of the slow wave \cite{Kim2024POP}. Spectroscopic measurements with SPRED \cite{Fonck1982AO} indicate that there is no influx of impurities associated with helicon power injection in DIII-D. Recent DIII-D experiments have demonstrated pitch angle scattering of runaway electrons with helicon power injection, a novel avenue for disruption mitigation research \cite{Choudhury2026PRL,Choudhury2026PPCF}. 

The DIII-D experiments presented here were conducted in L-mode plasmas with plasma currents of $\Ip = 0.85 - 1.1$ MA, on-axis toroidal field strengths of $\Bt = 1.9 - 2$ T, line-averaged electron densities of $\bar{n}_e \approx 2\ten{19}\mcubed$, and on-axis electron temperatures of $\Teo \approx 3 - 4$ keV. 
$\Bt$ and $\Ip$ were collinear, such that the edge magnetic field direction at the face of the antenna was within 4$\deg$ of the tilt of the antenna modules.
In addition to the helicon injection, the plasmas are heated with 2.3 MW of steady neutral beam injection (NBI) and $0.2 - 1.5$ MW of core electron cyclotron heating (ECH). No significant MHD activity is observed in these plasmas prior to the onset of sawteeth late in the shots, beyond which no analysis is performed. In these plasmas, the density is sufficiently high such that 476 MHz slow waves are evanescent for $\rho < 0.6$, guaranteeing that any observed heating and current drive close to the axis, where raytracing predicts helicon power absorption, is directly due to the launched helicon waves. 

\newcommand{\Xf}{\hat{P}}
\newcommand{\Yf}{\hat{T}_e}
\newcommand{\Pdep}{S}
\sectext{Power Deposition}
Helicon power deposition on electrons is a prerequisite for current drive and more straightforward to verify. Previous DIII-D helicon experiments had measured statistically significant signatures of electron heating in a few specific L-mode discharges \cite{Pinsker2024NF}. 
In this section, we present unambiguous evidence of electron heating due to helicon waves, which has been repeatedly observed in many DIII-D plasmas. An example from a typical L-mode is shown in \figref{fig:temod}. In this discharge, the klystron power was intentionally modulated at a frequency of 23 Hz in order to isolate the $\Te$ response. The core $\Te$ response is shown from a single channel of electron cyclotron emission (ECE), increasing when the helicon is on and falling in between pulses. Similarly clear evidence of electron heating due to the helicon is observed on many core-localized ECE channels, with the square coherence between the coupled power and $\Te$ response exceeding the 95\% statistical significance threshold, further described in the End Matter. 

The local power deposition can be estimated from the amplitude of the modulated $\Te$ response via an electron energy conservation equation, if transport is neglected: $S(\rho) \approx \frac{3\pi}{4}\omod\neo(\rho)\dte(\rho)$. Here, $\dte$ is the amplitude of the 90$\deg$ lagging response to the helicon power at the modulation frequency $\omod$, calculated via cross spectral analysis as  
$\dte \defined \smIm{\cpsd{\Xf}{\Yf}}\smabs{\Xf}/\cpsd{\Xf}{\Xf}$. 
Here, $\Xf$ and $\Yf$ are the Fourier images of the coupled helicon power and $\Te$, respectively. $\langle\dots\rangle$ denotes ensemble averaging of 50\% overlapping segments of the time series data, as further described in \citeref{Pinsker2024NF}. The blue curve in \figref{fig:powdep} shows the helicon power deposition profile calculated from the ECE array in this zero transport limit. The error bars indicate statistical uncertainty from the finite ensemble average used to calculate $\dte$ \cite{Bendat1978JSV,BendatText}. 

Due to the low $\betae$ in this L-mode plasma (1.1\% on-axis, 0.2\% volume averaged), the GENRAY ray tracing code \cite{Smirnov1995APS} predicts that approximately 40\% of the injected power will be absorbed on the first pass and localized near the axis. This represents a conservative estimate of the heating in case there are enhanced losses on multiple passes, which can not be modeled with certainty. The predicted power deposition profile is shown as the green curve on \figref{fig:powdep}, showing good agreement between the region of strongest absorption in the simulation and experiment. However, the absorbed power density predicted by GENRAY is significantly more peaked than the observed deposition profile calculated from the ECE measurements without accounting for transport. 

Since the helicon power was slowly modulated at 23 Hz, the $\Te$ response can be influenced by thermal transport on the same timescale as the helicon pulses, spreading out the apparent electron heating relative to where ray tracing predicts the power will be absorbed. In order to model the simultaneous electron response to the helicon injection and transport effects, time-dependent simulations are performed with TRANSP, a 1.5D tokamak power balance and transport code \cite{Hawryluk1980transp,Goldston1981JCP,Grierson2018FST,Pankin2025CPC,Pankin2026CPC}. For this analysis, TRANSP is coupled to GENRAY and also to turbulent transport models in order to predict $\Te$ over time, including the effect of transport, as the helicon source is modulated at the same frequency as in the experiment. Then, the TRANSP-predicted power deposition profile in the presence of transport is inferred from the predicted $\Te(\rho,t)$ evolution via the exact same calculation that was applied to the experimental data previously. 
The resulting profile is averaged across several simulations with different transport models in order to account for inherent model uncertainty, with additional details given in the End Matter. The result gives the orange curve in \figref{fig:powdep}, demonstrating a pronounced flattening of the inferred profile relative to the GENRAY prediction, in qualitative agreement with the experimental observations. Hence, the inclusion of transport effects explains the significantly less peaked power deposition profile observed in the experiment in comparison to the ray tracing prediction. 

In the L-mode regime of weak absorption, GENRAY predicts that the first pass absorption should become stronger with increasing $\betae$. To test this prediction, $\betae$ was varied in a series of discharges by changing the amount of ECH power used to heat the plasma, varying the background $\Te$. The green points in \figref{fig:betascan} show the first pass absorption predicted by GENRAY in each of these shots, increasing from 10\% to 45\% as the core $\betae$ is increased from 0.6\% to 1.1\% (volume averaged over $\rho < 0.2$). The blue points with error bars indicate the total measured electron heating due to the helicon in each shot, representing the integral of the power deposition profile in the absence of transport. As predicted, stronger absorption is observed as $\betae$ is increased across the shots, with the measurements tracking the GENRAY predictions. A more quantitative comparison would require correcting the observed power deposition for transport effects by fitting several harmonics of fluctuation data \cite{DeBoo2012POP,Petty2020AIP}, which is challenging due to the relatively low absorbed power $\like 100$ kW in these discharges. Previous analysis of modulated ECH has found that neglecting transport tends to underestimate absorption \cite{Petty2019APS}, suggesting that the transport-corrected helicon absorption may be somewhat higher than shown in \figref{fig:betascan}. Since the helicon power was modulated at the same frequency in each of the shots shown, the transport correction factor is expected to be comparable across the scan, such that the observed trend of stronger heating with increased $\betae$ is insensitive to the quantitative transport correction.  

\begin{figure}[tb]
\subfloat{\includegraphics[width = \columnwidth]{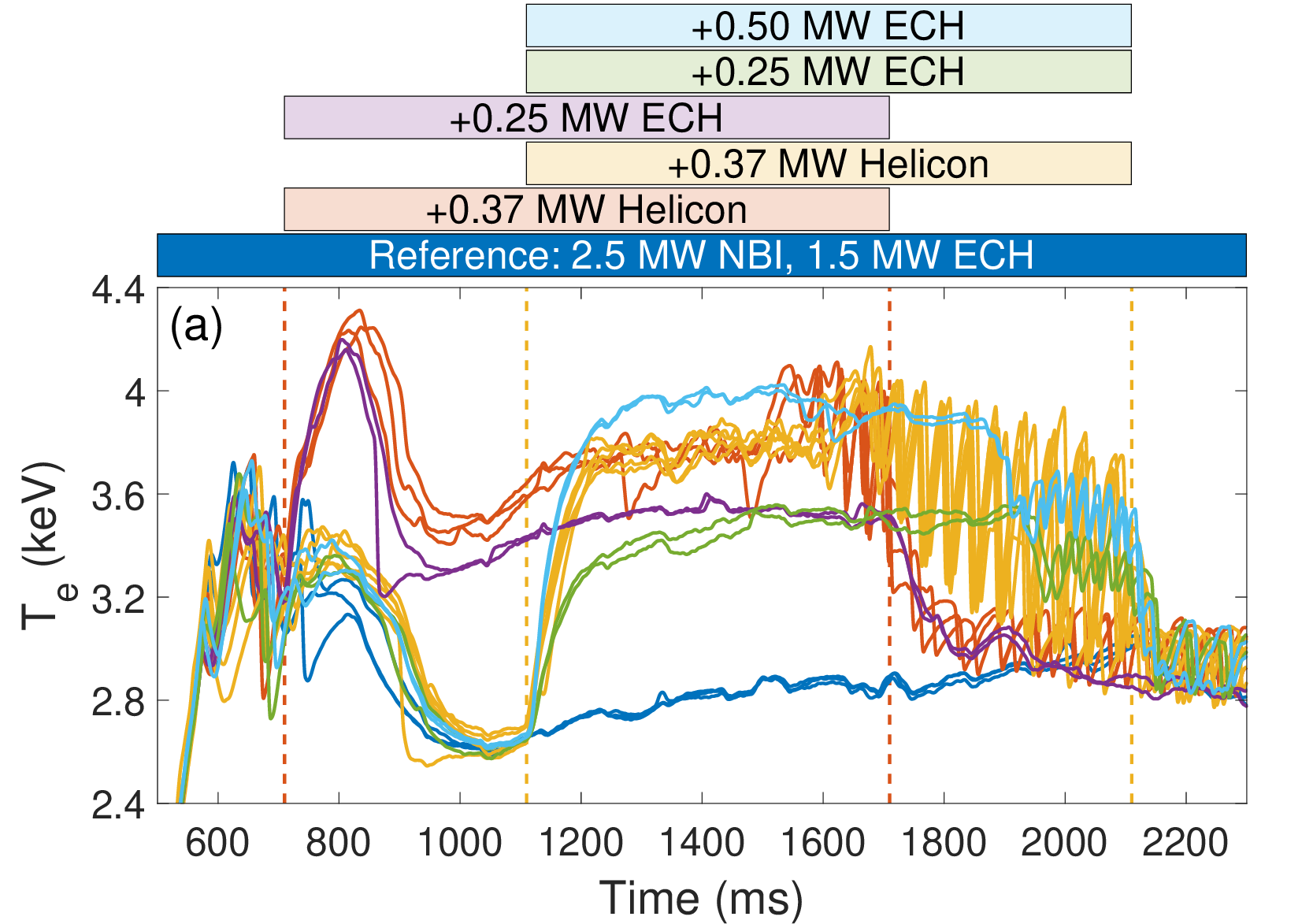}\label{fig:tetrace}} \\ \vshrink
\subfloat{\includegraphics[width = \columnwidth]{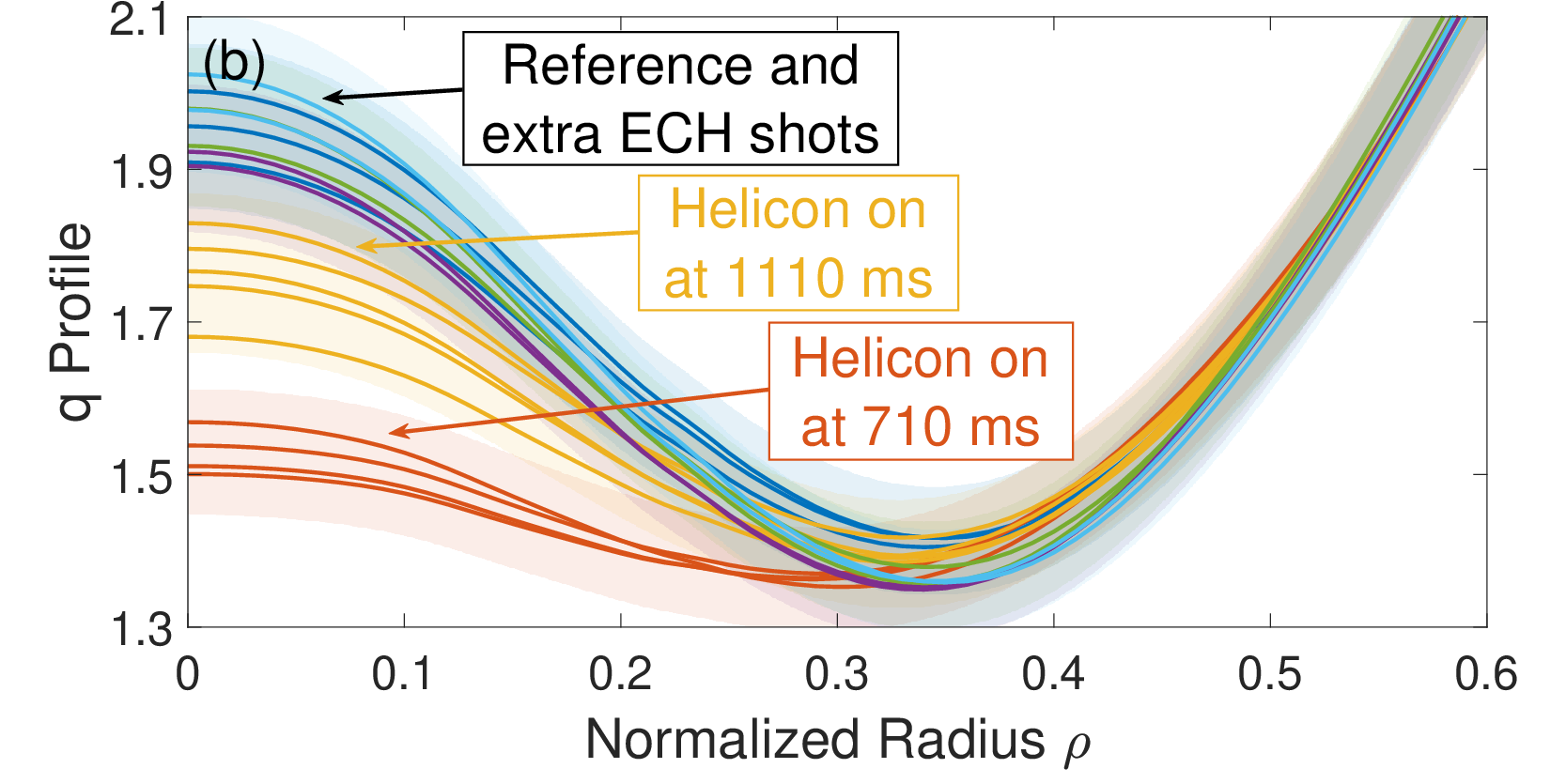}\label{fig:qprofile}} \vshrink
\caption{(a) $\Te$ evolution over time, with 10 ms smoothing, measured at $\rho \approx 0.02$ by ECE. (b) MSE-constrained EFIT reconstruction of the $q$ profile, averaged over $1300 - 1400$ ms. In both panels, each curve represents data from a separate shot, and each color represents shots with the same actuator waveforms. In (b), the shaded region represents the standard deviation from an average over shots, combined with a $5\%$ systematic MSE uncertainty \cite{Holcomb202607}. Data from a subset of DIII-D discharges 202141 -- 202161.}
\label{fig:evolution}
\end{figure}

\sectext{Evidence of Current Drive}
A dedicated L-mode experiment was performed in order to definitively demonstrate and quantify current drive from helicon waves. 
At the time of this experiment, only one side of the antenna was operational, precluding a direct comparison of co- versus counter-$\Ip$ current drive. 
A series of discharges were taken to distinguish between changes in the evolution of the current profile due to auxiliary current drive \vs the indirect effect of heating. Since current diffuses resistively, the current profile can be modified by changes in $\Te$, even if no additional non-inductive current is driven. To identify the current driven by helicon waves, three types of shots were compared: 1) shots with helicon power injected in the direction predicted to drive co-$\Ip$ current, 2) identical shots without helicon injection, and 3) identical shots where the helicon injection was replaced by an amount of ECH power comparable to the predicted first pass helicon absorption. These three types of shots represent the effect of 1) combined heating and current drive, 2) no additional heating or current drive, and 3) heating only, allowing the effect of helicon current drive to be isolated. 

The near-axis $\Te$ evolution, as measured by ECE, is shown in \figref{fig:tetrace} for all of the shots in this experiment. The dark blue curves represent the ``reference'' shots with the standard 2.3 MW NBI, 1.5 MW of ECH, and no helicon injection. The orange and gold traces are shots that are nominally identical to the reference shots, but with 370 kW coupled helicon power injected continuously without modulation from $710 - 1710$ ms (orange) or $1110 - 2110$ ms (gold). In both cases, the near-axis $\Te$ rises rapidly in response to the helicon injection, resulting in a $\like 1$ keV increase. GENRAY predicts 47\% first pass absorption for the shots with helicon injection, with a peaked deposition profile near the axis, similar to the one shown in \figref{fig:powdep} for a similar L-mode plasma. 

The purple, green, and light blue curves will be referred to as the ``ECH shots'', where the helicon power was replaced by a comparable amount of ECH with the same timing. The additional ECH was deposited near $\rho \approx 0.1$, the smallest radius accessible by the second harmonic (X2) ECH resonance at this field strength. Hence, the $\Te$ rise at $\rho = 0.02$ shown in \figref{fig:tetrace} should not be construed as a quantitative comparison of the total helicon \vs ECH absorption in these plasmas, but rather an indication of a similar amount of heating near the axis, where the helicon is most strongly absorbed. The several curves of each color on \figref{fig:tetrace} represent multiple separate shots taken with each distinct heating waveform, demonstrating robust reproducibility of the observations.  

Magnetic equilibrium reconstructions are performed with the EFIT code \cite{Lao1985NF,Lao2005FST}, with the current profile constrained by motional Stark effect (MSE) polarimetry \cite{Holcomb2008RSI}. \figref{fig:qprofile} shows the reconstructed safety factor ($q$) profile, averaged from $1300 - 1400$ ms, a time window when the helicon or replacement ECH are on in all shots, and before sawteeth begin. At this time, the reference and ECH shots all have similar $q$ profiles, with an on-axis value of $q_0 \approx 2$. Due to early NBI heating to delay the sawteeth, and the $\Ip$ ramp ending at 800 ms, the $q$ profiles are not yet fully relaxed. In the helicon shots, the reconstructed $q$ profile has flattened significantly, with more prominent flattening observed in shots where the helicon power injection began earlier in the shot (orange), consistent with the theoretical prediction that the helicon waves are driving co-$\Ip$ current on-axis. 

Additionally, the ECE data in \figref{fig:tetrace} shows that the helicon injection changes the sawtooth behavior. In the shots with helicon, the sawteeth first onset around $1500 - 1600$ ms, as evidenced by the large triangular oscillations in $\Te$. In contrast, the shots without helicon are either completely free of sawteeth (dark blue and purple curves) or have sawteeth starting much later in the discharge, as in the light blue and green curves, where the sawteeth begin around $1900 - 2000$ ms. Since sawteeth are triggered when the current profile first relaxes to $q = 1$, one can conclude that co-$\Ip$ helicon current is driving $q$ down faster than in the other shots. A similar modification of sawteeth behavior was previously observed on DIII-D due to FWCD in the ICRF \cite{Petty1997AIP,Petty2001PPCF}. This provides independent confirmation of helicon current drive, corroborating the evidence from MSE discussed above. 

Since the observed $q_0$ drops more rapidly and the sawteeth are triggered earlier in shots with helicon power than in those that replace the helicon with comparable or even greater ECH power, these observations can be confidently attributed to helicon current drive instead of byproducts of electron heating. Moreover, the plasma density in these shots was sufficiently high to prevent any incidentally excited or mode-converted slow waves from propagating to the core, excluding indirect LHCD as an alternative explanation.

\begin{figure}[tb]
\subfloat{\includegraphics[width = \columnwidth]{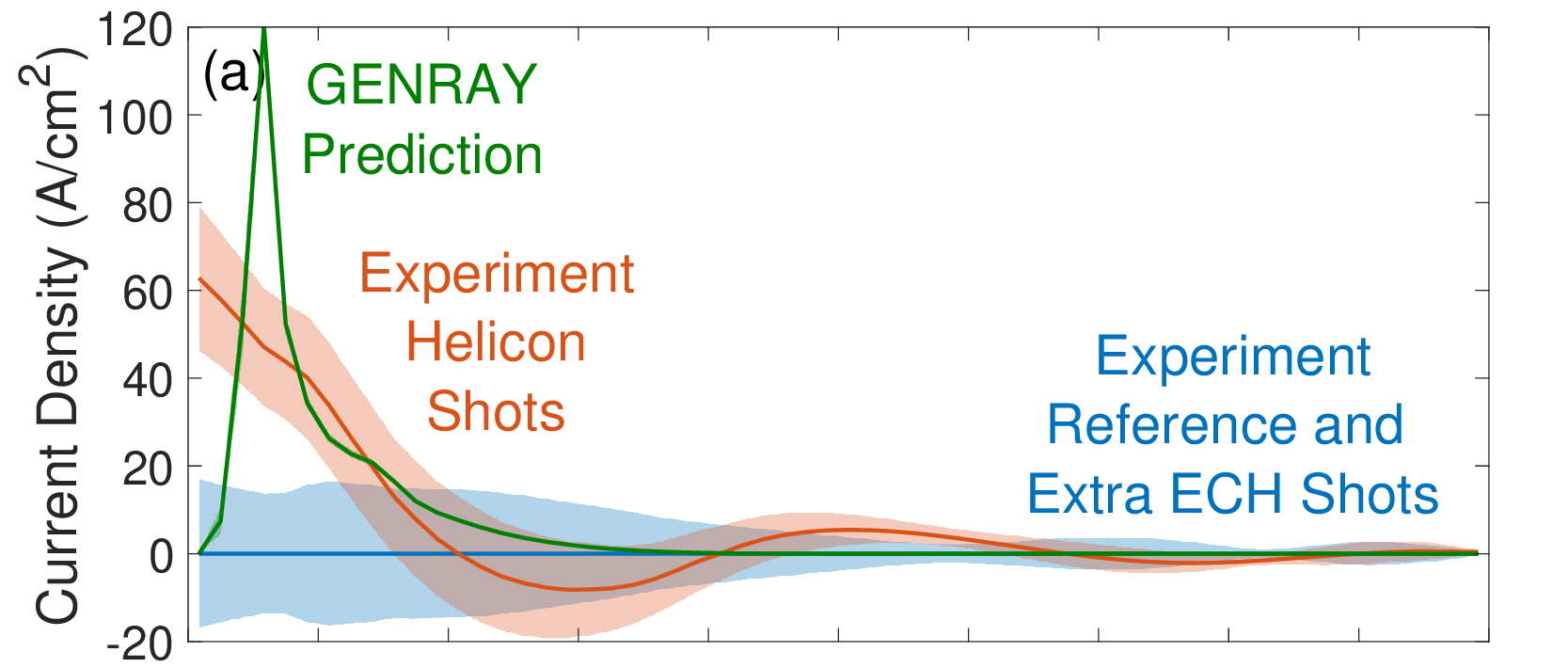}\label{fig:jni}} \\ \vshrink
\subfloat{\includegraphics[width = \columnwidth]{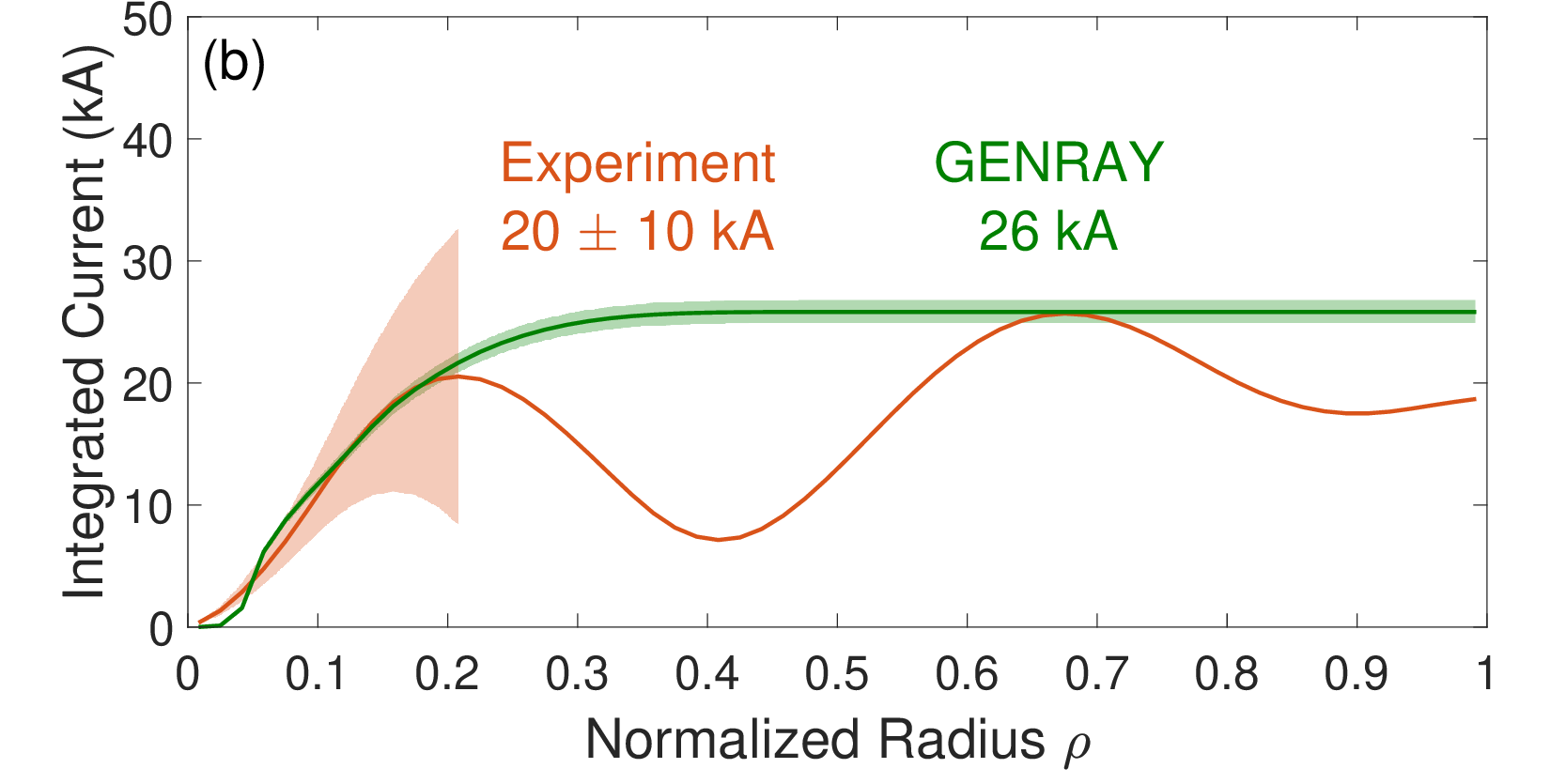}\label{fig:ini}} \vshrink
\caption{(a) Helicon current density profile and (b) radially-integrated helicon current. In both, shading represents the standard deviation from an average over shots. Orange curves show the experimental observations, green shows the GENRAY prediction for the first pass, and the shaded blue region in (a) shows the variation in reference shots without helicon injection.}
\label{fig:current}
\end{figure}

\newcommand{\jtot}{J_\parallel}
\newcommand{\jni}{J_\text{NI}}
\newcommand{\jhk}{J_\text{H}}
\newcommand{\ihk}{I_\text{H}}
\newcommand{\johm}{J_\text{Ohm}}
\newcommand{\jbs}{J_\text{BS}}
\newcommand{\jbeam}{J_\text{NB}}
\newcommand{\jec}{J_\text{EC}}
\newcommand{\epar}{E_\parallel}
\newcommand{\zeff}{Z_\text{eff}}
\sectext{Calculation of the Driven Current Profile}
To determine how much current was driven by helicon waves, it is necessary to calculate the remaining non-inductive current profile that can not be attributed to any other sources in the experiment. 
This approach was first described in \citeref{Forest1994PRL} and subsequently used to successfully interpret non-inductive current drive profiles in the DIII-D \cite{Forest1996POP,Forest1997PRL,Petty1999NF,Luce1999PRL,Petty2001PPCF,Murakami2003PRL,Chen2022NF}, JT-60U \cite{Naito2002PRL,Fujita2005PRL}, and NSTX \cite{Menard2006PRL,Gerhardt2011NFcur} tokamaks. By subtracting the Ohmic current profile and all ``known'' sources of non-inductive current  (\ie calculated by previously validated models), the helicon current drive profile is calculated as
$\jhk = \Delta \jtot - \Delta\johm - \Delta\left(\jbs + \jbeam + \jec\right)$. 
Here, $\Delta\jtot$ is the difference in the reconstructed parallel current profile in shots with and without helicon injection. $\Delta\johm$ and $\Delta(\jbs + \jbeam + \jec)$ are the same difference for the Ohmic current and the other non-inductive current sources, which in these plasmas were bootstrap, neutral beam, and electron cyclotron currents. Calculating the change in the non-inductive current between shots with and without helicon injection helps to mitigate systematic uncertainties in the analysis. The five shots with helicon power injected starting at 1110 ms (gold in \figref{fig:evolution}) will be averaged over and compared to the reference shots and all ECH shots with injection starting at 1110 ms (dark blue, light green, and light blue in \figref{fig:evolution}, seven shots total). Analysis is performed from $1360 - 1560$ ms, the longest quiescent window available prior to the onset of sawteeth. The Ohmic current profile is calculated as $\johm = \sigma \Epar$, where $\sigma \propto \Te^{3/2}/\zeff$ is the neoclassical conductivity \cite{Sauter1999POP} evaluated using kinetic profile data and $\Epar \propto \partial\psi/\partial t$ is computed from EFIT reconstruction. The ``known'' non-inductive currents are calculated by models implemented in TRANSP, as detailed in the End Matter.

The helicon current profile resulting from this calculation is given by the orange curve in \figref{fig:jni}, with the shaded region representing the standard deviation from averaging over shots. The current profile is peaked in the core and concentrated within $\rho < 0.2$, similar to the heating profile measured via power modulation in similar L-mode plasmas (\figref{fig:powdep}). Moreover, the measured helicon current drive near the axis far exceeds the spread in the residual non-inductive current density in shots without helicon injection (shaded blue region), demonstrating that this observation can not be explained by random variation in the plasma conditions or systematic uncertainty. The corresponding GENRAY prediction for the current driven on the first pass is shown in green, with the current localized near the axis, as measured in the experiment. Although the current profile predicted by GENRAY is more peaked than the observation, this peak value is highly sensitive to the simulated ray trajectories, as it depends on how close the rays pass to the magnetic axis, where the cross sectional area rapidly vanishes. 

The radially-integrated helicon current is shown in \figref{fig:ini}. Strong agreement is found between the measured and predicted integrated current up until $\rho \approx 0.2$, where radial oscillations in the observed current density result in significant variation in the integrated current beyond this point. Although these oscillations, an artifact of the EFIT reconstruction, are much smaller than the on-axis current density, they nonetheless have a nontrivial influence on the integral due to the increasing plasma area with $\rho$. Hence, the total driven current implied by integrating all the way to the boundary has significant uncertainty. A more reliable quantity for this analysis is to integrate the current up to $\rho \approx 0.2$, where the calculated current density first changes sign, and beyond which the oscillations dominate the calculated current density and are comparable to the variation across the reference shots. This yields a driven helicon current of $20 \pm 10$ kA, reported to one significant figure due to integration uncertainty on the order of the unphysical oscillations. By comparison, GENRAY predicts 21 kA of driven current enclosed in $\rho < 0.2$ and 26 kA total on the first pass of the plasma. 

Since GENRAY predicts about 50\% of the power is absorbed on the first pass, this translates into a GENRAY prediction of at most $\approx 50$ kA driven, if there was complete multipass absorption of the 370 kW of power coupled to this plasma with absolutely no losses. The present experimental analysis can not distinguish whether the measured $20$ kA of current drive is due to incomplete absorption or a reduced current drive efficiency. This requires more rigorous quantification of the power deposition, which is ongoing and beyond the scope of this Letter, which is to present the first definitive evidence of helicon current drive. 

\sectext{Conclusions}
Core electron heating has been directly observed in DIII-D plasmas, providing clear evidence of power deposition by helicon waves launched with a traveling wave antenna. The measured absorption is consistent with time-dependent integrated modeling including the effects of transport, and scales with $\betae$ as predicted. Helicon current drive has been demonstrated for the first time in any device. The reproducible evidence includes changes in MSE-constrained EFIT reconstruction and sawtooth timing that can not be explained by heating alone. Calculation of the non-inductive current profile found good agreement with ray tracing predictions. These results represent experimental proof-of-principle of helicon current drive, which theoretically extrapolates to efficient, off-axis current drive in reactor-grade plasmas. 

\sectext{Acknowledgments}
The authors thank M.E. Austin and N.J. Richner for fruitful discussions. The TRANSP simulations reported here were performed with computing resources at the Princeton Plasma Physics Lab. Part of the data analysis was performed using the OMFIT integrated modeling framework \cite{Meneghini2015NF}. This material is based upon work supported by the U.S. Department of Energy, Office of Science, Office of Fusion Energy Sciences, using the DIII-D National Fusion Facility, a DOE Office of Science user facility, under Awards DE-FC02-04ER54698, DE-SC0026408, and DE-SC0016154.

\sectext{Disclaimer}
\gadisclaimer

\onecolumngrid
\medskip
\begin{center} 
	{\Large\bf End Matter}
\end{center}
\twocolumngrid

\begin{figure}[tb]
\includegraphics[width = \columnwidth]{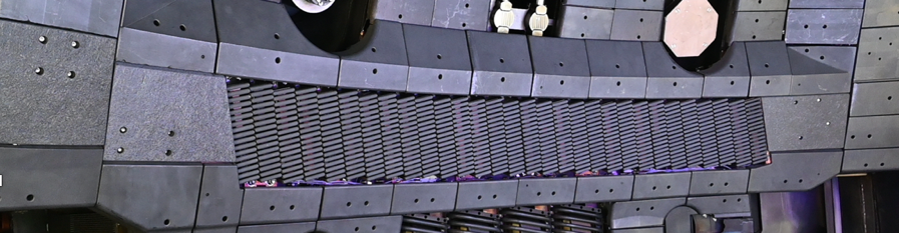}
\caption{The DIII-D helicon traveling wave antenna mounted on the outer wall. The toroidal direction is horizontal.}
\label{fig:twa}
\end{figure}

\sectext{Antenna Hardware}
The DIII-D helicon traveling wave antenna is shown in-vessel in \figref{fig:twa}. Not visible are radio frequency current probes embedded into 16 of the antenna modules. The helicon power coupled to the plasma is determined by fitting the exponential power decay through this probe array as power flows through the antenna toroidally. Greater than $90\%$ of the power reaching the antenna was routinely coupled to the plasma. Further details about the helicon hardware can be found in \citeref{VanCompernolle2021NF,Pinsker2024NF}. 

\sectext{Plasma Diagnostics}
The main diagnostics used to characterize the plasma are as follows. Thomson scattering provides $\neo$ and $\Te$ profile measurements \cite{PonceMarquez2010RSI}. An electron cyclotron emission (ECE) radiometer also measures $\Te$, with faster time resolution and better spatial resolution near the magnetic axis than Thomson scattering \cite{Austin2003RSI}. A motional Stark effect polarimeter (MSE) is used to make internal measurements of the pitch angle of the magnetic field \cite{Holcomb2006RSI}, with 13 channels covering up to $\rho \approx -0.3$ on the HFS and $\rho \approx 0.7$ on the LFS of the plasmas analyzed in this work. Charge exchange recombination spectroscopy (CER) was used to measure carbon impurity density and temperature profiles (in order to infer the main ion deuterium density and temperature) \cite{Burrell2004RSI} by blipping a 0.8 MW diagnostic beam with a 10\% duty cycle. 

\sectext{GENRAY Ray Tracing}
Two representative GENRAY ray tracing results for the plasmas analyzed in this work are shown in \figref{fig:genray}. The left panel shows discharge 200743, the same shot discussed in \figref{fig:heating} when analyzing the helicon power deposition, with $\Ip$ and $\Bt$ both counter-clockwise when viewed from above. The right panel shows discharge 202158, one of the several shots averaged over in \figref{fig:current} for calculating the helicon-drive current profile, with $\Ip$ and $\Bt$ both clockwise. In both simulations, 32 rays are launched from 35$\deg$ above the midplane, sampling a power spectrum $P(\npar) \propto \exp(-(\npar - 3)^2/0.3^2)$. The spread in ray trajectories is due to varying amounts of refraction for rays launched with different values of $\npar$. 

\newcommand{\rayheight}{5.5cm}
\begin{figure}[tb]
\hspace*{-0.2cm}
\subfloat{\includegraphics[height = \rayheight]{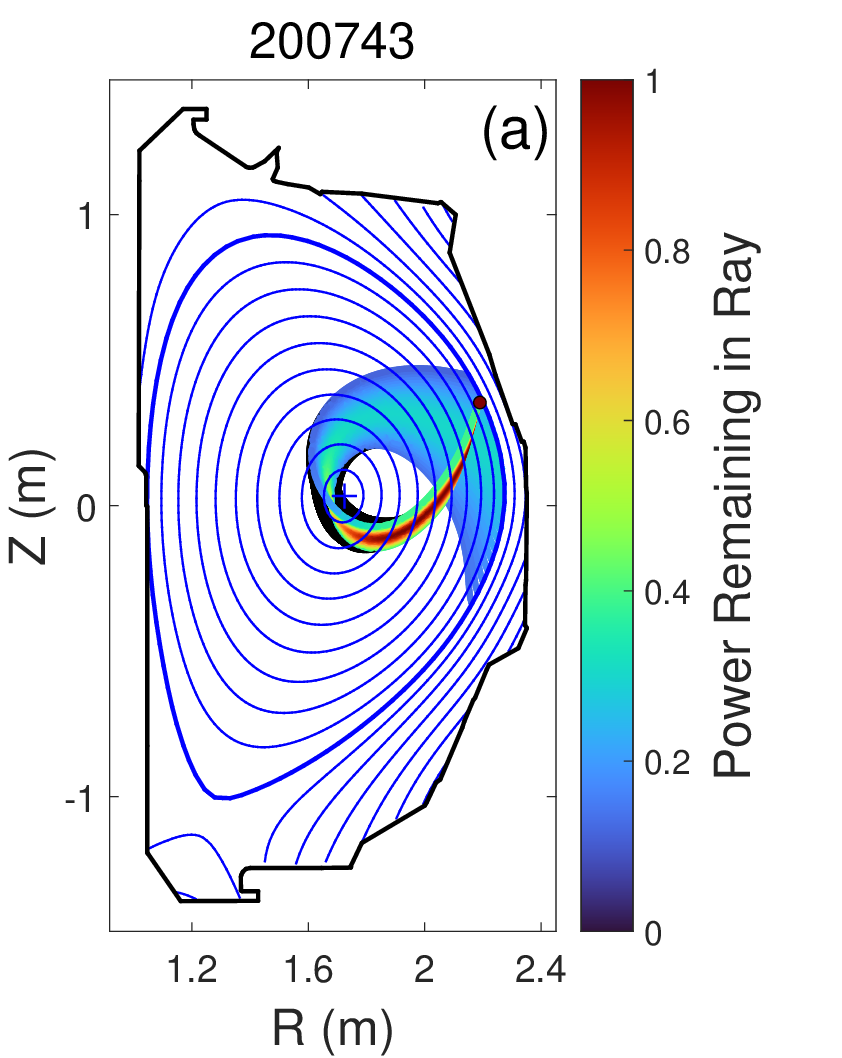}\label{fig:rays200743}} \hspace*{-0.2cm}
\subfloat{\includegraphics[height = \rayheight]{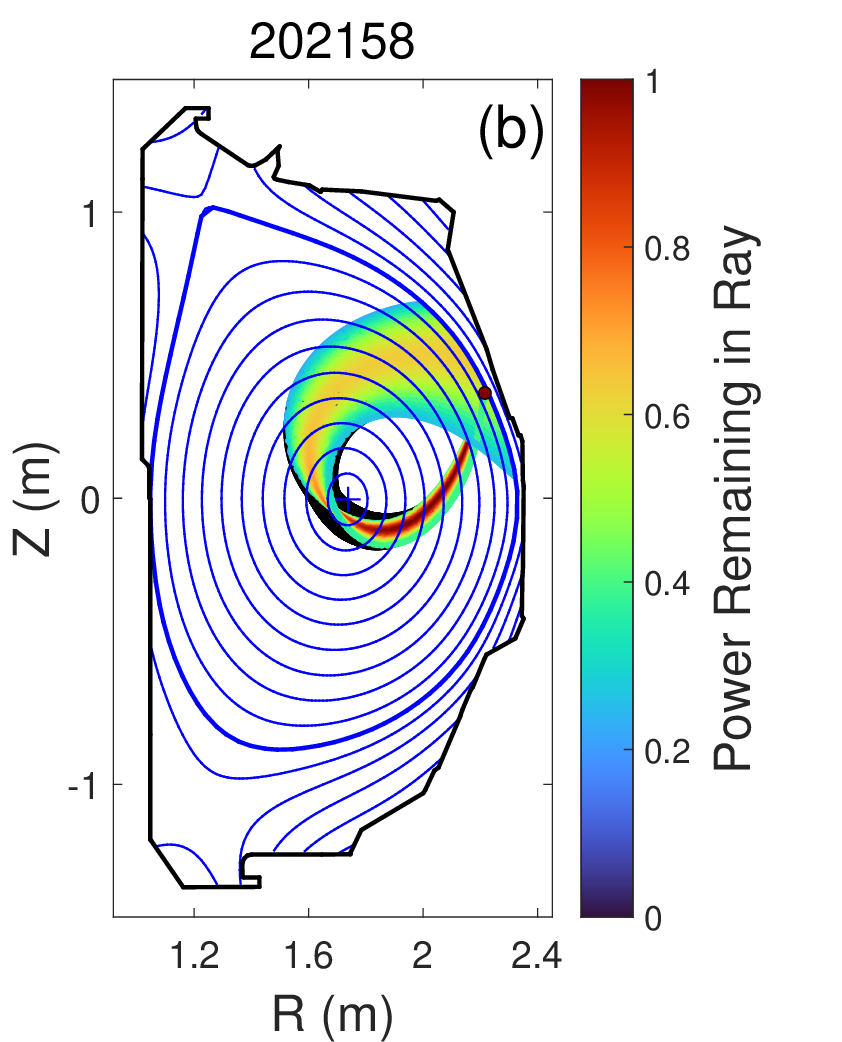}\label{fig:rays202148}} \hfill \\ \vshrink
\caption{GENRAY ray tracing results for example discharges used to analyze (a) power deposition and (b) current drive. The rays start in the simulation at the filled red circle. Colorscale is normalized by the injected power. The black shading indicates absorption on electrons.}
\label{fig:genray}
\end{figure}

\begin{figure}[tb]
\subfloat{\label{fig:cohfreq}}
\subfloat{\label{fig:cohrho}}
\includegraphics[width = \columnwidth]{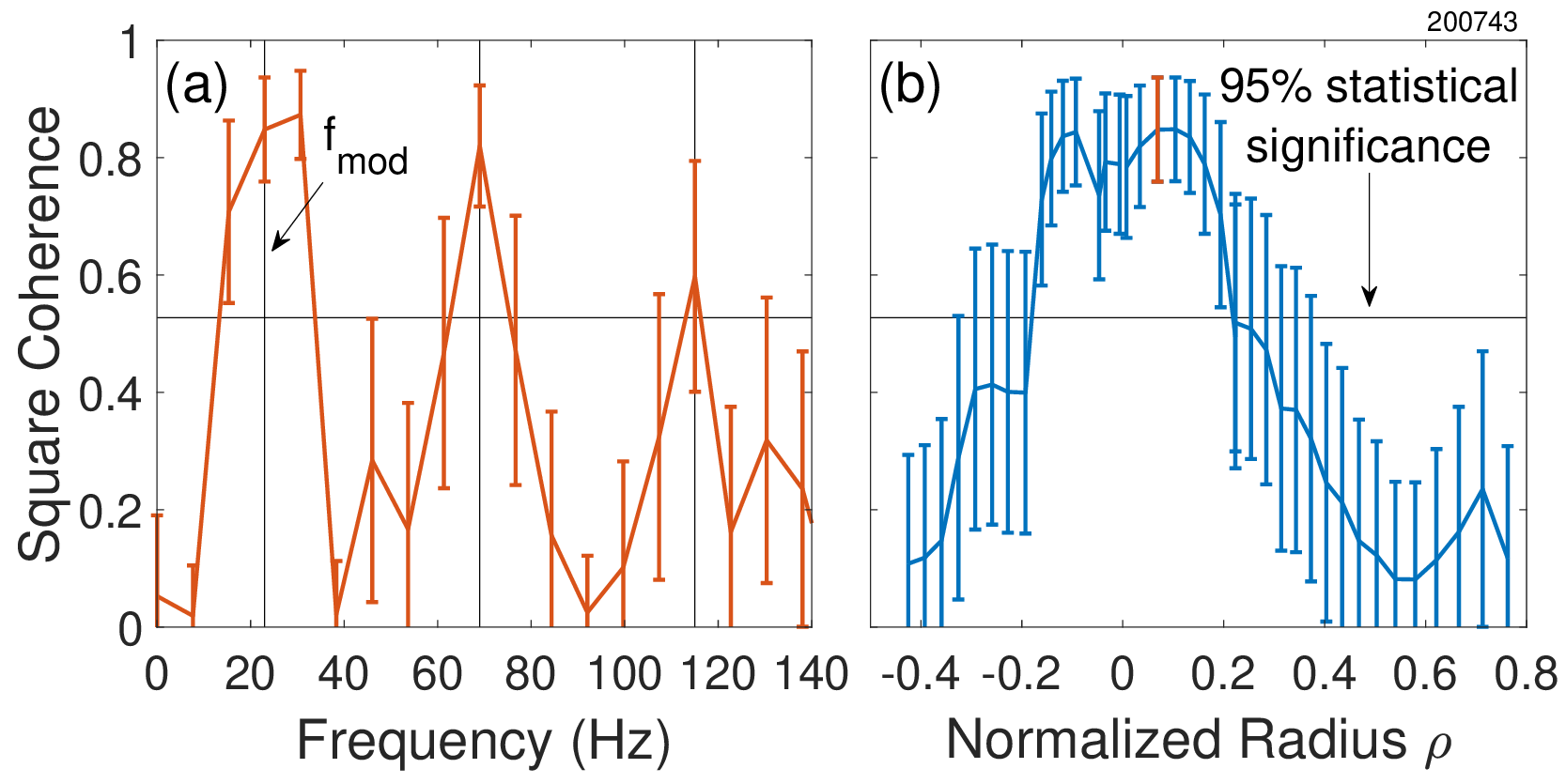} \\ \vshrink
\caption{Square coherence between injected helicon power and $\Te$. (a) Frequency dependence of $\gxy$ for a single ECE channel near $\rho \approx 0.07$. Vertical lines indicate odd harmonics of the 23 Hz power modulation. (b) Spatial dependence of $\gxy$ evaluated at the modulation frequency on the entire ECE array. On both, the horizontal line indicates the threshold for 95\% statistical significance. The orange point in (b) indicates the ECE channel plotted in (a).}
\label{fig:coherence}
\end{figure}

\sectext{Statistical Significance}
The square coherence can be used to quantify how well synchronized the electron temperature response is to the modulated helicon power over time. For each ECE channel, it is calculated as $\gxy(f) = \smabs{\cpsd{\Xf}{\Yf}}^2/\cpsd{\Xf}{\Xf}\cpsd{\Yf}{\Yf}$, where $0 < \gxy < 1$ by construction. For a given frequency, a value near one indicates that the phase of $\Yf$ tracks the phase of $\Xf$ closely in time, and a value near zero indicates a lack of correlation. The statistical significance level of a given value of $\gxy$ depends on the number of unique time segments used in the ensemble averaging procedure. For $n_s$ segments with $50\%$ overlap and defining $n = n_s/2$, the square coherence must be greater than $\gxy > 1 - \alpha^\frac{1}{n - 1}$ in order to reject the null hypothesis with $\alpha$ uncertainty. For instance, using $n_s = 9$ segments spanning $920 - 1600$ ms in discharge 200743, a 95\% confidence level $(\alpha = 0.05)$ corresponds to $\gxy > 0.53$. The square coherence for this shot is shown in \figref{fig:coherence}. \figref{fig:cohfreq} shows $\gxy(f)$ for the ECE data shown in \figref{fig:temod} near the axis, demonstrating prominent peaks at the 23 Hz modulation frequency and its odd harmonics. Even harmonics are absent since the square waveform used for the modulated helicon power only contains odd harmonics. \figref{fig:cohrho} shows $\gxy(\rho)$ by evaluating the square coherence of the injected helicon power with the response from each of the 39 ECE channels at the modulation frequency. Several channels in the core far exceed the 95\% statistical significance level, indicating a strong, coherent electron temperature response near the axis, as predicted by ray tracing calculations for this plasma. 

\sectext{Predictive TRANSP Simulations} 
The predictive TRANSP simulations shown in \figref{fig:powdep} use the implicit transport solver \ptsolver to time evolve temperature profiles by solving an energy conservation equation for each desired species \cite{Pankin2025CPC,Pankin2026arxiv}. On each time step, \ptsolver uses transport coefficients that are calculated by turbulent transport models (\ie not inferred from power balance analysis of experimental inputs). In this work, simulations are performed with both the Multi-Mode Model (MMM) \cite{Rafiq2013POP,Luo2013CPC} and the Trapped Gyro-Landau Fluid (TGLF) code \cite{Staebler2007POP,Kinsey2008POP}. A neural net surrogate model of TGLF (TGLFNN) \cite{Neiser2024APS,Lestz2025PPCFa} is used in order to reduce the computational cost of resolving the modulated helicon power in time. MMM consists of four different submodels, each one treating a different class of instabilities. TGLF is a drift wave model that employs physics-based saturation rules that have been calibrated to nonlinear gyrokinetic simulations in order to calculate quasilinear fluxes. TGLF saturation rules SAT1 and SAT3 are used in this work, motivated by the findings of a large DIII-D database validation of TGLFNN within a time slice flux-matching solver \cite{Neiser2024APS}. 

For the DIII-D discharge shown in \figref{fig:powdep}, six different predictive TRANSP simulations are performed. MMM, TGLFNN SAT1, and TGLFNN SAT3 are each used in simulations that predict $\Te$ alone or $\Te$ and $\Ti$ simultaneously. Since TRANSP time evolves the entire temperature profile, not just the perturbative response to the helicon waves, the modulated heating source that is calculated by GENRAY is sensitive to the background $\Te$ profile that each model predicts. \eg if one transport model predicts the background $\Te$ to be lower than the fitted experimental profiles, GENRAY would calculate less absorption in that simulation than using the experimental profiles, precluding a meaningful comparison of the time-dependent temperature response, since the modulated source would be weaker than expected for the experimental profiles. In order to control for this, the injected helicon power was scaled in each simulation such that the GENRAY absorption in each predictive simulation was within $3\%$ of the GENRAY absorption calculated for the experimental profiles. By employing this matching procedure, the six distinct predictive TRANSP simulations with different turbulent transport models are all predicting an electron temperature response to the same modulated heat source, such that the range of $\Te(\rho,t)$ responses that they predict can be attributed to differences in the underlying transport models. Hence the error bars on the orange curve in \figref{fig:powdep} represent the combined uncertainty of the statistical error from the finite ensemble averaging procedure that is applied to the synthetic $\Te(\rho,t)$ data and also the transport uncertainty that is captured by averaging over several different models. 

\begin{figure}
\subfloat{\label{fig:jtot}}
\subfloat{\label{fig:johm}}
\includegraphics[width = \columnwidth]{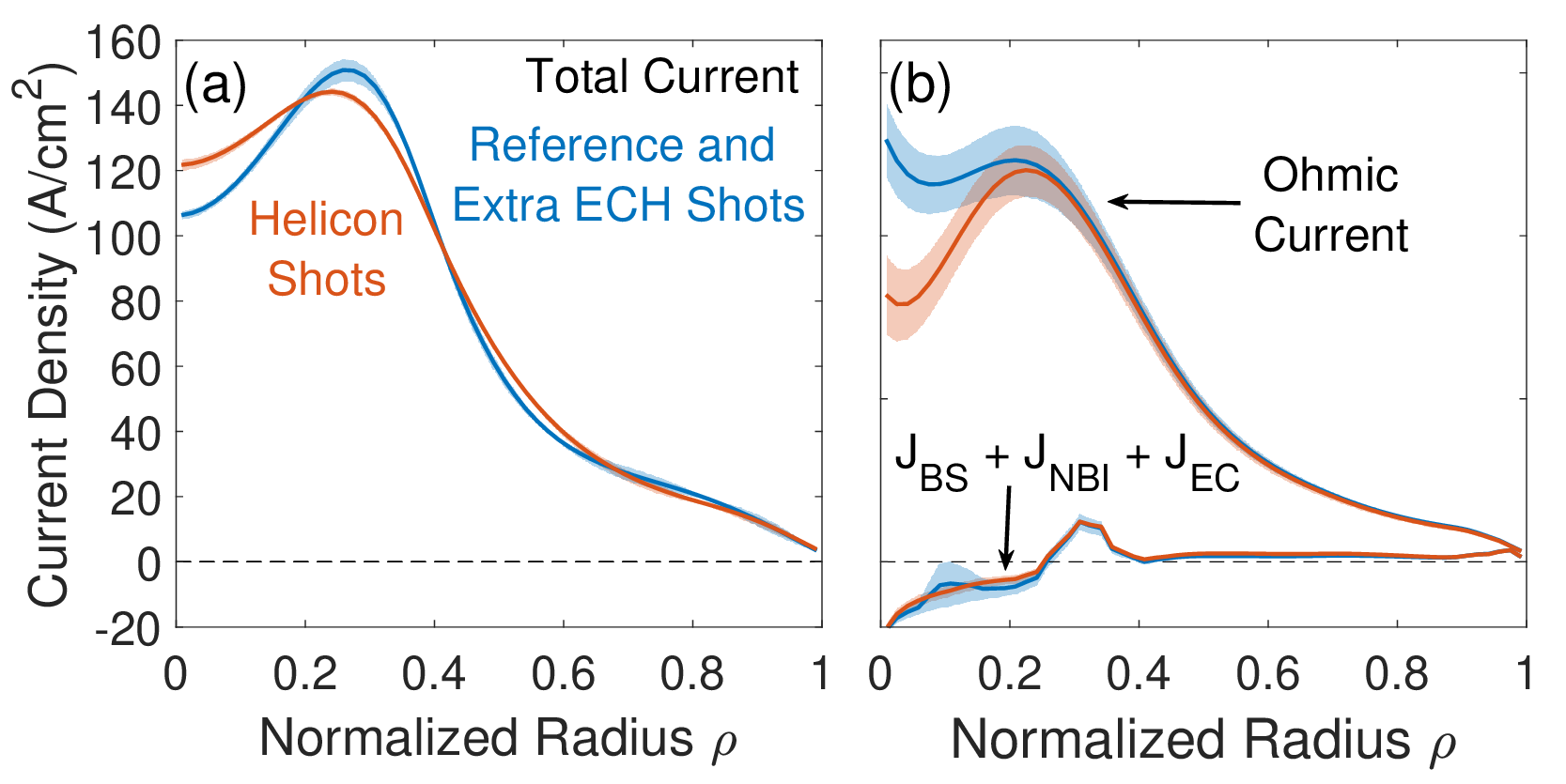} \\ \vshrink
\caption{Current profiles used in the calculation of the driven helicon current. (a) Total parallel current, calculated by MSE-constrained EFITs. (b) Ohmic current profile derived from EFITs and kinetic profile measurements, alongside other sources of non-inductive current present in these discharges, calculated by models in TRANSP: bootstrap ($J_\text{BS}$), neutral beam $(J_\text{NBI})$, and electron cyclotron $(J_\text{EC})$ currents.}
\label{fig:jother}
\end{figure}

\sectext{Other Current Sources}
The driven helicon current profile shown in \figref{fig:jni} results from subtracting all other sources of current from the total parallel current calculated via magnetic equilibrium reconstruction. For completeness, these current profiles are shown in \figref{fig:jother}. The parallel current from MSE-constrained EFITs is given in \figref{fig:jtot}, where the systematic differences near the axis between the shots with and without helicon power injection correspond to the changes in the $q$ profile shown in \figref{fig:qprofile}, which are calculated from the same EFITs. \figref{fig:johm} shows both the derived Ohmic current profile and the numerically modeled non-inductive current sources. The non-inductive currents are calculated in TRANSP using the Sauter model for $\jbs$ \cite{Sauter1999POP}, the Monte Carlo NUBEAM code for $\jbeam$ \cite{Pankin2004CPC}, and the TORAY-GA ray tracing code for $\jec$ \cite{Matsuda1989IEEE}. In this L-mode plasma, the bootstrap current is negligible. The neutral beam current is negative since the beams were injected in the direction opposite to the plasma current. The small peak in the non-inductive current near $\rho \approx 0.3$ is a small amount of residual ECCD from the nearly radial launch of electron cyclotron waves used to heat the background plasma. The sum of the beam, neutral beam, and electron cyclotron current sources being nearly equal in the shots with and without helicon injection is a consequence of the discharges being carefully matched. To account for systematic uncertainty in the $\zeff$ profile in this experiment, a flat $\zeff$ profile was used in all calculations, averaging over a range of $\zeff = 2 - 2.4$ for each shot, consistent with the best available CER data. Due to this variation in $\zeff$ having a similar effect on the Ohmic current calculated in shots with and without helicon injection, it contributes a systematic uncertainty of $\pm 1$ kA on the inferred helicon current drive, much smaller than the total $\pm 10$ kA uncertainty due to all factors.

\bibliography{all_bib} 

\end{document}